\documentclass{sig-alternate}
\usepackage{graphicx}
\usepackage{epstopdf}
\usepackage{wrapfig}
\usepackage{moreverb}
\usepackage{flushend}

\begin{document}

\title{A Hands-on Education Program on Cyber Physical Systems for High School Students}

\numberofauthors{3}
\author{
\alignauthor
Vijay Gadepally\titlenote{Vijay Gadepally is a PhD candidate at The Ohio State University and corresponding author}\\
       \affaddr{Ohio Supercomputer Center}\\
       \affaddr{1224 Kinnear Road}\\
       \affaddr{Columbus, Ohio}\\
       \email{vijayg@osc.edu}
\alignauthor
Ashok Krishnamurthy
       \affaddr{Ohio Supercomputer Center}\\
       \affaddr{1224 Kinnear Road}\\
       \affaddr{Columbus, Ohio}\\
       \email{ashok@osc.edu}
\alignauthor 
Umit Ozguner\\
       \affaddr{The Ohio State University}\\
       \affaddr{Electrical and Computer Engineering}\\
       \email{umit@ece.osu.edu}
}
\date{30 July 1999}

\maketitle
\begin{abstract}

Cyber Physical Systems (CPS) are the conjoining of an entities' physical and computational elements. The development of a typical CPS system follows a sequence from conceptual modeling, testing in simulated (virtual) worlds, testing in controlled (possibly laboratory) environments and finally deployment. Throughout each (repeatable) stage, the behavior of the physical entities, the sensing and situation assessment, and the computation and control options have to be understood and carefully represented through abstraction.

The CPS Group at the Ohio State University, as part of an NSF funded CPS project on ``Autonomous Driving in Mixed Environments'', has been developing CPS related educational activities at the K-12, undergraduate and graduate levels. The aim of these educational activities is to train students in the principles and design issues in CPS and to broaden the participation in science and engineering. The project team has a strong commitment to impact STEM education across the entire K-20 community.

In this paper, we focus on the K-12 community and present a two-week Summer Program for high school juniors and seniors that introduces them to the principles of CPS design and walks them through several of the design steps. We also provide an online repository that aids CPS researchers in providing a similar educational experience.

\end{abstract}




\section{Introduction}
\label{intro} 

As the Cyber-Physical Systems Group (CPS) at The Ohio State University, under an NSF funded project entitled ``Autonomous Driving in Mixed Environments'', we have been planning educational activities to promote student interest in the Science, Technology and Mathematics (STEM) fields. The need for promoting STEM related education to middle-high school students in the STEM fields has been widely documented \cite{perezhappens,tyson2007science,kuenzi2006science} and can be summarized as highlighting the need for the United States to prepare a sufficient number of STEM professionals capable of innovation. As described in \cite{lowell2009steady}, STEM education can be visualized as a pipeline that begins in early education, extends through college and ends with employment with critical transition points that includes the high school to college transition.

The CPS group at Ohio State has significant experience working designing K-12 education related activities (in addition to college undergraduate and graduate student courses). In \cite{sivilotti:2003q}, the authors discuss a project designed for middle school girls as a part of a week-long workshop, entitled ``Future Engineers' Summer Camp'' held at The Ohio State University. The authors describe their approach to introducing middle school girls to fault tolerant computing through a variety of kinesthetic learning activities. Kinesthetic learning activities \cite{sivilotti:2007q}, are proposed by the authors as a process by which students learn about theoretical concepts by carrying out physical activities, as opposed to passively listening to lectures. The authors describe their success using kinesthetic learning activities to explain complex algorithms, such as sorting, to younger audiences. In \cite{sivilotti:2010u}, the authors discuss coursework developed at the undergraduate level to teach students the important concept of abstraction.  

In this paper, we present a two-week educational program for high school students that introduces them to Cyber Physical Systems, and the design and development of such systems through the modeling and simulation at differing levels of abstraction. Students are introduced to CPS through the engineering of \textit{autonomous vehicles} or \textit{driverless cars}. In particular, students are asked to develop algorithms for a vehicle (in this case, a Roomba) that can avoid obstacles. The aim of the program is to help students emulate the scientific process employed by CPS researchers while learning to use common techniques and tools. This paper aims to describe the activities and provide sufficient resources to facilitate the emulation of such a program.

CPS design, due to the cost and difficulty of direct physical testing, typically goes through four phases in system design:

\begin{enumerate}
\item \textbf{Conceptual Modeling}: Understand the mathematics of the problem, and propose a theoretical solution (develop equations of motion, develop analytical solutions where possible.)
\item \textbf{Simulated Testing}: Use computer simulations to validate algorithms (use a software package with a simulated test bed to test obstacle avoidance algorithms.)
\item \textbf{Controlled Environment Testing}: Use a physical test-bed of a simulated environment to validate algorithms (use the developed algorithm on a physical Roomba within a simulated environment.)
\item \textbf{Real World Deployment}: Test the algorithm in the real world (use the developed algorithm on a physical vehicle on actual city streets.)
\end{enumerate}

The activities described in this paper are conducted as a part of the Summer Institute (SI) held at the Ohio Supercomputer Center (OSC). SI is a two-week residential program for gifted highschool freshman, sophomores, and juniors designed to raise students' interest and awareness of the STEM fields. SI challenges students to cultivate their research ability through the use of cutting-edge tools, modeling and simulation, and interaction with active researchers. Students are also encouraged to develop interpersonal skills through presentations and participation in a variety of science related field trips, and teambuilding activities. 

\begin{figure}
\centering
\epsfig{file=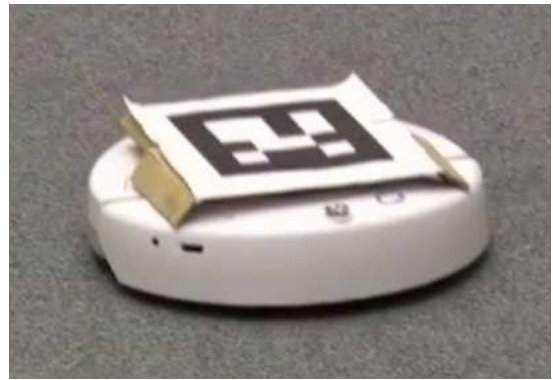, scale=0.5}
\caption{Roomba fitted with GPS tag}
\label{roombaPicture}
\end{figure}

The CPS project developed for this purpose is called ``Obstacle Avoidance Roombas,'' and is a direct product of autonomous vehicle research carried out by CPS researchers at Ohio State. Over two years, 2010 and 2011, eight motivated students were chosen to participate in the project. These students were introduced to real-life aspects of CPS designs, trained in C/C++ programming, and taught relevant mathematics and physics concepts. Students were then asked to use a simulator, called Player/Stage, to program (in C/C++) Roombas (as shown in Figure~\ref{roombaPicture}) to complete the project. The project was divided into a sequence of four subproblems to help students understand the logical project progression.

\begin{enumerate}
\item Program a Roomba to follow a set of coordinates entered by a user
\item Program a Roomba to acquire a target, and plan the optimum path to reach the target
\item Program a Roomba to acquire a target, and avoid a single obstacle to reach the target
\item Program a Roomba to acquire a target, and avoid multiple obstacles to reach the target
\end{enumerate}

Students are taught how simulations can provide a path to real world implementation, and use developed code on a set of robots and obstacles in a laboratory setting at Ohio State University. 

In this article, we present details of this project, the educational materials developed, and results obtained. We begin by giving an overview of the SI program, the CPS related project, and the intended competencies. We present the logical progression of the project through the four step process taken by CPS researchers. We conclude the project description with student feedback and lessons learned by the CPS staff. In order to facilitate similar projects, we end with a list of the resources that were used in developing this project.


\section{About OSC's Summer Institute}

For over 20 years, the Ohio Supercomputer Center (located in Columbus,OH) has offered the Summer Institute (SI) to Ohio's gifted students entering their sophomore, junior or senior years of high school and their teachers. SI is a two-week residential program designed to raise students' interest in the Science, Technology, Engineering and Mathematics (STEM) fields through a collaborative and dynamic research environment and hands on experience with the latest in cutting edge technology. The program is held in Columbus, OH and students live in the dormitories of The Ohio State University for this two-week period. Students typically arrive at OSC by 9 AM every morning, and work on their chosen projects until 5 PM when they are taken back to their dorm rooms and take part in a variety of social activities. Each year, a number of projects are chosen by SI staff that appeal to students. Projects are decided taking into account previous student feedback, real-world applicability of project, staff expertise and funding. 

The program begins by teaching students UNIX, the operating system of the computers they use. Next, students learn a programming language (C/C++/MATLAB) and any software required to complete their projects. Students are required to do their own work from code implementation to final presentations. The ability to develop algorithms and an understanding of the project's science/engineering basis are needed. 

\subsection{SI 2010 and 2011}

Thirty two students, and four teachers participated in the SI's held in 2010 and 2011. In SI 2010 and 2011, there were four projects ranging from robotics to medical imaging. The four projects with descriptions (as given to students) are given in \cite{si-project}. The project described in this paper, Obstacle Avoidance Roombas, was presented to SI participants as follows:

\vspace{6pt}

\noindent\textbf{Obstacle Avoidance Roombas}: \\

\textit{Many organizations, such as the US Army, require vehicles that are capable of avoiding dangerous objects. These objects may be explosive obstacles, booby trapped buildings, or armed personnel. This sponsored project involves students using robots and robotic simulators to design an ``Obstacle Avoidance Roomba.''}

\textit{The project uses a Roomba, an autonomous robotic vacuum cleaner, fitted with LIDAR (a light based ranging device). The goal of the project is to program the Roombas to avoid randomly placed obstacles to reach a destination. Students work with a vehicle simulator, called Player/Stage, to design ``the brains'' of a Roomba . After successfully simulating the behavior, the research group will do ``real-life'' tests of the Roomba in a specially fitted laboratory.}
\\

The SI program began with a presentation of four projects and students were allowed to choose one based on their interests.

\subsection{Project: Obstacle Avoidance Roombas}
\label{problem}

The project involves the design of algorithms and code that directs a Roomba to avoid obstacles and reach a target or goal. Initial programming and testing of code is done using the Player/Stage~\cite{player} simulator. Player is a network server for robot control that supports a variety of robot hardware. Stage simulates (in 2.5D) a population of mobile robots, sensors and objects. The Player/Stage package allows quick prototyping of algorithms for implementing embedded computers. The Player/Stage environment is designed to simulate a set of roads and intersections set up in the Control and Transportation Laboratory  (CTL) testbed, at The Ohio State University. Sensors used in the testbed are simulated in the Player/Stage environment. Thus, the students see how a "real-world" laboratory environment can be abstracted and modeled in a simulated environment. The necessity of a ``cyber'' environment is further demonstrated by the ease with which code prototyped in the simulator can be transferred to testbed equipment. Students are also introduced to limitations of the simulated world, and situations in which the simulator may allow a violation of physics or mathematical possibilities in the real world such as a simulator valid Roomba position that translates to a Roomba hovering above the ground that cannot be attained physically. 

In general, the two weeks of SI are divided into training and project components. The first week consists primarily of providing students with  tools and any mathematical or physical foundations required to complete their projects. Students also make multiple laboratory visits to gain an understanding of the environment that they are simulating.

\subsection{Competencies}
\label{competencies}

The CPS research team decided on various competencies that would need to be taught to students. 

\subsubsection{Mathematics} 

Students were given a two hour interactive lecture that introduced them to the mathematical concepts required to complete Obstacle Avoidance Roomba project. The lecture began with a refresher on coordinate geometry and covered concepts such as: frames of reference, coordinate and homogeneous transformations. Student competency was tested with simple mathematical problems such as: ``The roomba is facing 45$\,^{\circ}$ in the Roomba frame, what would be the corresponding angle in the Earth frame?''

\subsubsection{Physics}

Students were given a guest lecture about the physics behind CPS fundamentals. Students were taught that vehicles are often modeled as a point-mass or a bicycle to simplify calculations. Students were then taught about the point-mass model and Bicycle model of vehicles. Students also reviewed Newton's laws, friction, and simple dynamics. Student competency was tested informally through questions and answers.

\subsubsection{Tools}

\begin{figure}
\begin{center}
\includegraphics[scale=0.7]{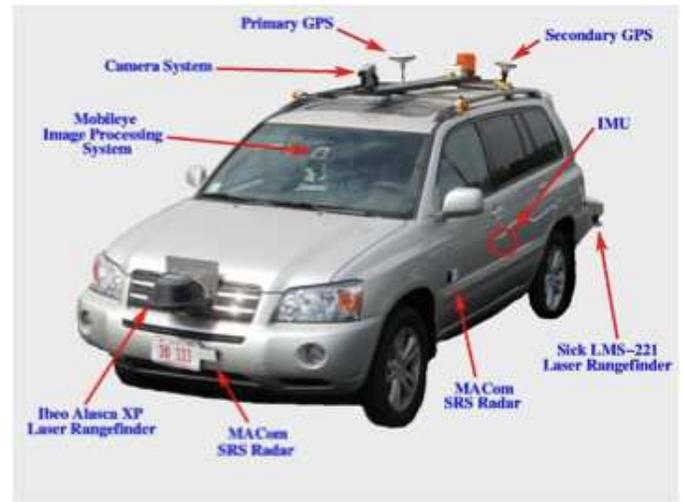}
\end{center}
\caption{OSU ACT Vehicle and Sensors}
\label{vehicle}
\end{figure}

Students were introduced to various tools used by CPS researchers in the design and deployment of autonomous vehicles. Students were given a tour of Ohio State's Center for Automotive Research and one of the autonomous vehicles as shown in Figure~\ref{vehicle}. Instruction concentrated on the sensors used in the vehicle, namely the GPS systems, Laser Rangefinder and Radar systems. Student competency was tested informally through questions and answers.

\subsubsection{Programming}

Students were introduced to programming of different languages used in CPS design. Instruction concentrated on C/C++ (two popular programming languages) and MATLAB (a very high level programming language popular with engineers). Students were given a two hour lecture on programming basics, that concentrated on syntax, and commands. Training materials was taken largely from previously developed training material available at \cite{training}.  Student competency was tested throughout the project.

\subsubsection{Scientific Process}

The scientific process employed by CPS researchers, described in Section~\ref{intro}, was the central theme of the Obstacle Avoidance Roomba project. Students were reminded throughout the project of the process and that the logical progression of the project followed this process.


\section{Phase 1: Conceptual Modeling}

In the first phase, students are provided with details of the problem to be solved and asked to develop algorithms for each sub-part of the problem. The goal of this phase is to introduce students to the process of learning the mathematical and physical competencies described in section~\ref{competencies} and turning them into a proposed algorithm. Another goal is to introduce students to the concept of peer review.

Students are also given real life examples of autonomous vehicles performing obstacle avoidance. Student also learned about the sensors used in Roombas, giving them an understanding of the type of data that they can use for developing their algorithms. Student learning is tested continuously through simple exercises.

Once students are comfortable with the competencies required for project completion, they are split into two groups of two student each. Students then work in their sub-group collaboratively develop high level algorithms for the four parts of the problem. In order to simulate the peer-review process of scientific development, students are asked to present their algorithms for each of the problem parts to the other group and instructor for comments and feedback.


\section{Phase 2: Simulated Testing}

Once students have developed algorithms that pass the peer review process, they are asked to use computer simulations to validate their algorithms for each of the sub-problems. The goal of this phase is to provide students with hand-on experience with the simulation tools (one of the intended competencies) used by CPS researchers, in addition to introducing students to the process of simulated validation of algorithms.

\subsection{Training and Tools}

Students use the Player/Stage package, a commonly used robot simulation program. Player provides a network interface for a variety of robot hardware, such as the Roomba. Stage is a mobile robot simulator that provides handles to a variety of sensor models. The environment used within Player/Stage has been developed by the Center for Intelligent Transportation Research (CITR), and simulates the physical testbed, called SimVille, available for laboratory testing at CITR. SimVille is described later in this paper. The Player/Stage program uses the C/C++ programming language to control the simulated Roombas movements. Developed code can be transferred directly to the actual Roombas in the physical testbed.

To introduce students to the Player/Stage syntax, students are walked through the solution of the first problem. Students are then asked to convert the algorithms developed in Phase 1 into Player/Stage compatible C/C++ code.

\subsection{Activities}

Students spent approximately twenty hours programming in C/C++ with the Player/Stage environment. Their goal was to solve the remaining three parts of the problem as mentioned in section~\ref{problem}. At the end of each problem, students are asked to present their code (to emulate the code review process), and simulations to other students and the project lead (for peer review). 

\begin{figure}
\epsfig{file=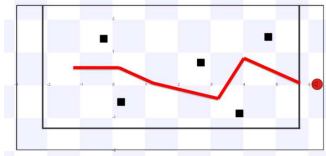, scale=0.35}
\caption{Simulated Version of Third Part of Project (Red Circle represents the Roomba, Black Squares Represent Obstacles)}
\label{lab3_simulated}
\end{figure}

In order to promote the development of robust algorithms and code, students try to provide conditions that may ``break'' the code or algorithm. Further, the instructor may modify the environment, such as moving the obstacle. The aim of this task is to teach students the fundamentals of designing robust code, and the fact that research and development is an iterative process. Once students and instructor are satisfied with the robustness of code and algorithm, they proceed to the next part of the problem. An example of the simulated output for Problem 4 is given in Figure~\ref{lab3_simulated}. For this part, students use Player/Stage to program the Roomba (red circle) to pass through all the Obstacles (black squares).

\subsection{Difference in Student Approaches}

Students were encouraged to use the the math and physics they were taught to develop their own algorithm for controlling the Roomba. For example, in the third part of the project, where students were asked to program a Roomba to reach a target by avoiding a single obstacle, one group approached the problem by coding a set of switch statements that would move the Roomba in a deterministic manner depending on which quadrant of the screen the obstacle is in. 

\begin{figure}
\epsfig{file=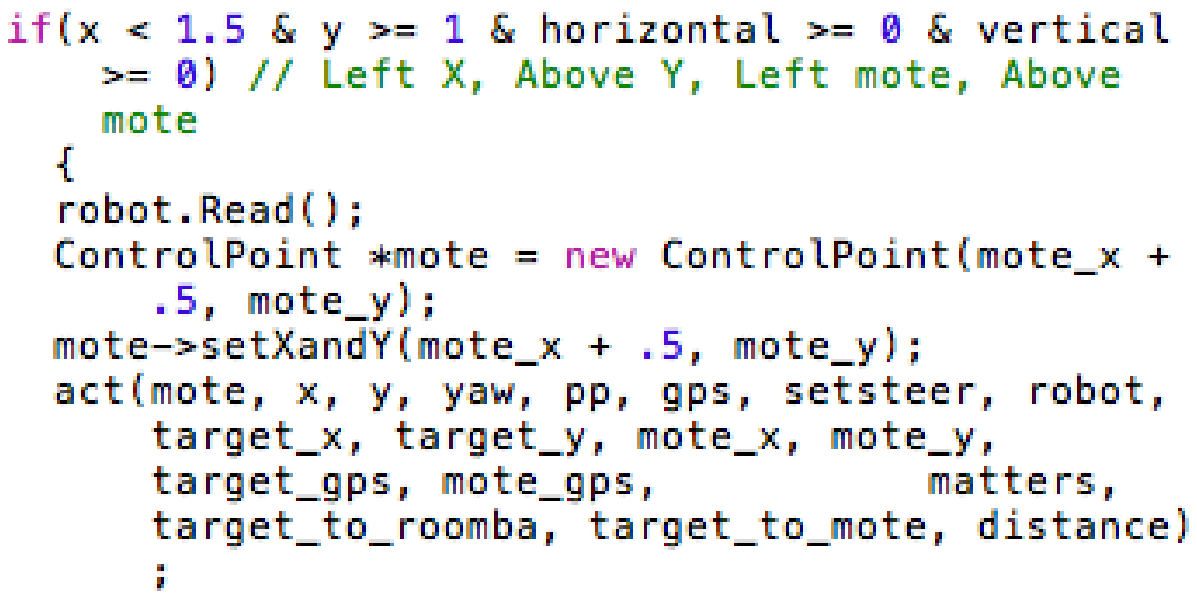, scale=0.64}
\caption{Code Sample of Group 1}
\label{code1}
\end{figure}

\begin{figure}
\epsfig{file=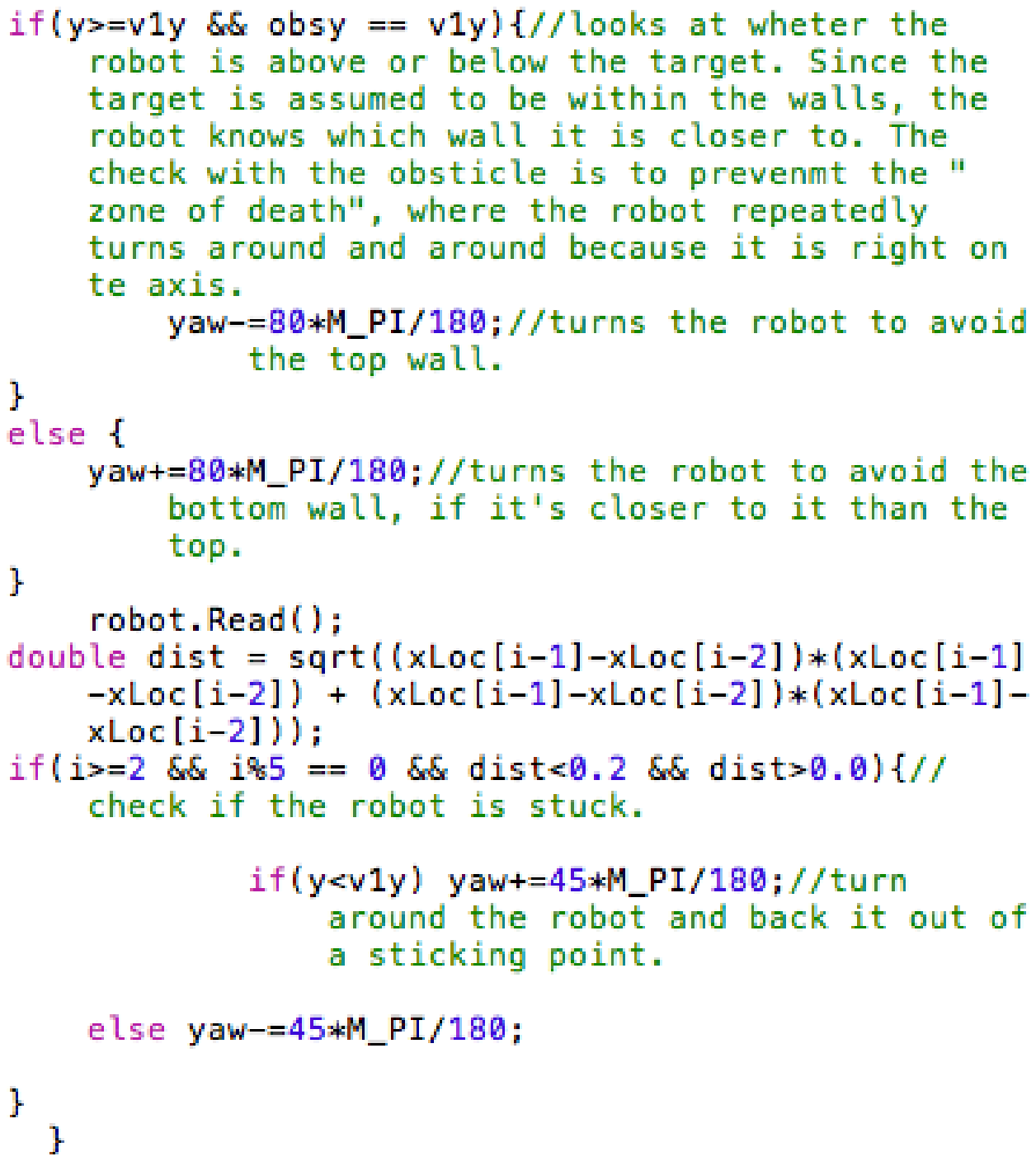, scale=0.64}
\caption{Code Sample of Group 2}
\label{code2}
\end{figure}

Another group programmed the Roomba to travel to a fixed point away from the obstacle. The two code samples of Figures~\ref{code1} and~\ref{code2} show one such difference in approach.

\section{Phase 3: Controlled Environment Testing}

\begin{figure}
\begin{center}
\includegraphics[scale=0.3]{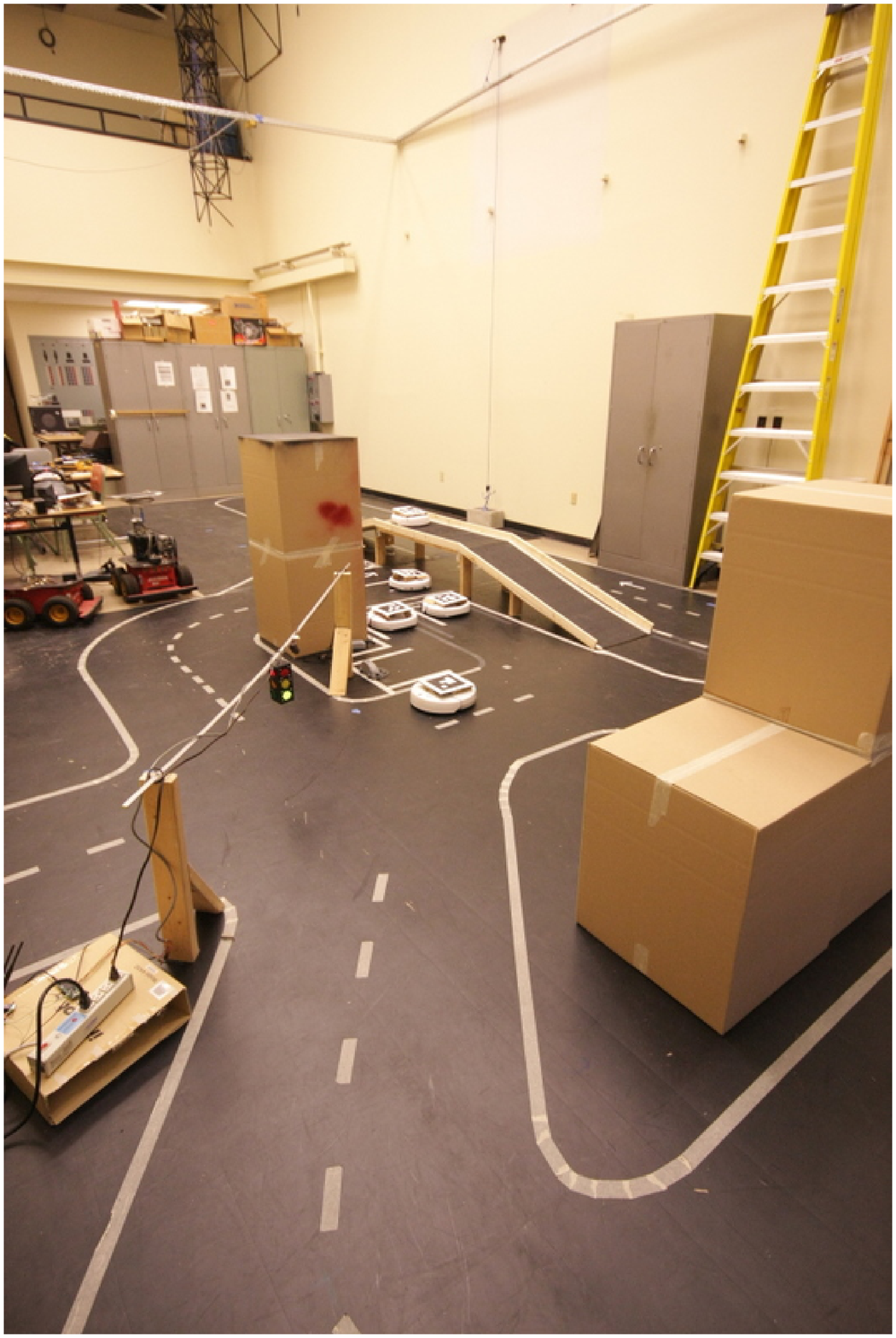}
\end{center}
\caption{ Testbed}
\label{testbed}
\end{figure}

Once students have sufficiently robust code, they are asked to use a controlled representation (laboratory setting) of the real world - a testbed, called SimVille. SimVille was created in 2007 in order to expedite research efforts in urban environment scenarios. SimVille \cite{vernier2010virtual} is a 1/7 scale road network that is designed to provide easy access to a road network. Additionally ceiling mounted cameras provide a ``Virtual GPS'' system that robots (Roombas) using SimVille can get information about their location, location of obstacles, etc. The Player program works as the network interface to communicate between Roombas, GPS sensors, and control code. Control code is a slightly modified version of code written for the Stage simulator. A sample configuration of SimVille is given in Figure~\ref{testbed}. The goal of this phase is to introduce students to the concept of a controlled (laboratory) environment, and illustrate the parallels between code developed for the simulated environment and code needed for the testbed environment. Students also learn the difference between these settings. For example, some of the Roomba speed values used in the simulated environment cannot be physically realized in the testbed environment because of physical limitations of the Roomba motors, and floor material which are approximated in the simulator.

\subsection{Presentation and Tools}

Students are given multiple tours of the testbed, over the course of the algorithm development and simulator proof-of-concept phases, to give them an understanding of the lab capabilities, and physical constraints. Additionally, students are allowed to \textit{play} with the roombas to understand the dynamics of movement. Students are shown how to translate code written in Stage to the actual Roombas, through a varied compilation technique. Student are also given a detailed view of the sensors present in the testbed.

Once students have programmed their algorithms for the four problems, they are asked to implement their code on the testbed. Students are able to modify the code they have written for Stage and use their algorithms on the actual Roombas. 

\subsection{Activities}

While students are testing their code with the Roombas in SimVille, SI staff introduce perturbations to the environment by moving obstacles to give students an understanding of working with moving obstacles. Students are asked to modify their code, if necessary, to solve any issues that are brought about by these perturbations. Students complete their testing for all four parts of the problem by the end of the second week


\section{Phase 4: Real World Deployment}

While deploying the developed code on an actual vehicle is not feasible, given the short time span, students are introduced to this important phase through tours of several laboratories, including the Center for Automotive Research (CAR), which houses the OSU Autonomous Vehicle. Students also have an opportunity to visit other laboratories at The Ohio State University such as: a Bio-dynamics Laboratory, and a Virtual Reality Laboratory. The goals of these laboratory visits is to show students the practical aspects of their work. The goal of this phase is to show students the final phase of the CPS research process - real world deployment. 


\section{Closing and Feedback}

At the end of the two-week period, students are asked to present their research, and results to all of the SI students, Ohio State University faculty, SI Staff, and parents. Students are also asked to give their feedback, and any suggestions for future SI programs.

\subsection{Participant Feedback}

Upon completion of the Summer Institute program in 2010 and 2011, students were asked for their feedback of the project, and suggestions for how to improve the project in subsequent years. Excerpts from their comments are given below:

\subsubsection{Student 1}
``\textit{For the past two weeks, I have been enrolled in the Obstacle Avoidance Roomba project at the OSC Summer Institute. I have found it both informative and entertaining, and strongly urge that it be offered again next year. Learning C and the rudiments of autonomous-vehicle programming is engaging, and I enjoy the relaxed and informal working environment. However, I felt that the competitive arrangement of two pairs of programmers against one another was somewhat counterproductive: we probably would have been able to accomplish considerably more if we had pooled all of our resources.}''
Video feedback can be found at \cite{tomVideo}.

\subsubsection{Student 2}
``\textit{At first I was really just looking forward to an easy project with roombas since I've used them before. Not using any other sensors was a bummer, but I realize now that it'd be impossible to incorporate the sensors in our time frame. I liked working in a big group better than in our teams of 2, because it got really competitive at times. My favorite parts of the project were when the program actually worked when compiled and driving the roombas at Dreese labs. I've actually learned a lot about psuedocode, high level coding, and different approaches to obstacle avoidance and using the GPS sensors. Overall I really enjoyed working on the project, and it's definitely the highlight of my summer.}''

\subsubsection{Student 3}
``\textit{During the past two weeks I had lots of fun learning about Roombas, C Programming, and Quake 3. I thought that the pace of the project was slow enough that I did not feel rushed, yet fast enough to allow us to be productive. I think that it might have been interesting to replace the multi-obstacle lab with a lab having to do with sensors. We talked a lot about the importance of sensors, so I was a bit disappointed to discover that we would not be using them in our project. The camp was extremely fun because of all of the participating students and staff. This was probably one of the best two weeks I have had in a while.}''

\subsubsection{Student 4}
``\textit{When I was first assigned this project, my head was intrigued by the possibilities of what we could program the Roomba to do. Shortly after we started programming, however, we ran face to face to the difficulties of using C to tell the Roomba where and how to move. Project Obstacle Avoidance Roomba is an great assignment to enlighten those unfamiliar with Autonomous Vehicles. The Roomba Project illustrates the challenges of avoiding the walls, moving around the obstacles, even turning the Roomba. Although I had no part in writing the functions of controlling the Roomba, there was plenty of work of just coding in an algorithm the robot can follow to drive itself. Obstacle Avoidance Roomba Project would be a great stepping-stone to help interested newcomers step into the field of Autonomous Driving.}''

\subsection{Lessons and Future Projects}

The SI Staff learned many valuable lessons from students, and will use these to provide improved iterations of the project in subsequent years.

\subsubsection{Scope}

Students particularly enjoy the coding components of the project. The scope of the project is sufficient for students to have an understanding of CPS fundamentals without excessive training. Coding training given in C/C++ and is intended to help students with the Player/Stage programming. Basic mathematical training in coordinate transformations, homogeneous transformations in addition to basic physics is also provided to give students further understanding of the Roombas. Students expressed interest in using a larger set of sensors (as opposed to purely GPS coordinates). Inclusion of further sensors may be difficult within the 2 week length of the program.

\subsubsection{Competition}
Students at times felt that the competitive process was counterproductive and that they would rather that component not be in the next iteration of the project. Students indicated a preference for group based work - something they felt would be more productive. This modification was made in SI 2011 and groups worked in a collaborative manner while maintaining the peer review process. Students were also allowed to switch groups based on approaches or areas of interest. The collaborative group structure was more successful than the competing group structure.

\subsubsection{Working with varying student capabilities}
One of the difficulties faced by the project teams was instructing students with differing technical capabilities, especially in knowledge of programming. In SI 2010 students were paired such that each group would have one student proficient in programming working alongside a student who was not as familiar with programming. With student feedback that they felt this process required them to ``carry'' another student, in SI 2011, an additional instructor (who was also a SI 2010 student) was brought in to give personalized attention to students who required technical help - so as to ensure proficient students were not slowed down. 

\subsubsection{Metrics to Judge Project}
One metric used to judge program success in SI 2010 and 2011 is student willingness to participate in future robotics related activities after the Summer Institute program. Another metric, student learning of competencies, was tested through informal systems. In future iterations, the authors wish to devise a formal evaluation to judge student competencies.

\subsubsection{Project Evaluation}
In SI 2010 and 2011, evaluation was collected through a daily online journal that asked students the following questions: 

\begin{enumerate}
\item What did you learn today?
\item Who did you help out today and how?
\item Who helped you out today and how?
\item What did you like best about today's activities?
\item What did you like least about today's activities?
\end{enumerate}

Such information was collected over eight days culminating in a final comprehensive survey about the overall experience. Data was also collected about student-instructor interactions and improvements that could be made to the overall program and particular project. In the final survey, 100\% of the students ``Strongly Agree'' that the instructors were helpful. Multiple students cited that the ``High Point'' of the experience was in the Roomba testbed. One student cited a ``Low Point'' when they had difficulty with the coding component of the project. Overall, 75\% of the students felt that the programming portion of the project was enjoyable with 25\% feeling that they needed greater prior programming experience.  Students were also asked to comment about the experience such as lab tours, residence halls, etc. As a measure of being taught in a way that corresponds to learning style, 36\% said ``Strongly Agree'', 43\% said ``Agree'' and 21\% said ``Neutral.'' 100\% of the students said that the project has deepend their desire to work in the field of robotics or engineering. 


\section{Tools and Resources}

The aim of this paper is to present a CPS related study that can be recreated. This section outlines the tools/material used and basic instructions on creating a similar project. 

\begin{itemize}
\item  The first step is to download and install the Player/Stage project from http://playerstage.sourceforge.net. Download and installation instructions for a variety of hardware configurations is included in the instruction manual.
\item The programmable roombas used in the testbed component of the project can be purchased from http://store.irobot.com. The Player/Stage simulator can be configured to work with this Roomba.
\item To view and download the SI 2011 Source Code: \\http://dl.dropbox.com/u/1268613/codesamples.zip
\item To view and/or download the SI 2010 Student Presentation and Videos:\\ http://dl.dropbox.com/u/1268613/RoombaProject.zip
\end{itemize}

Other related links and resources:

\begin{enumerate}
\item Student Feedback Video: \\http://www.youtube.com/watch?v=Ke8ONfF-Q64
\item NSF CPS Program: \\http://www.nsf.gov/funding/pgm\_summ.jsp?pims\_id=503286
\item C/C++ Training Material: \\ http://www.osc.edu/supercomputing/training/
\end{enumerate}

\section{Conclusions}

We present a two week educational program for High School students as a part of the Summer Institute program at the Ohio Supercomputer Center. Students are introduced to CPS related fundamentals, and develop the algorithm and code for an obstacle avoidance Roomba. Students are taught the scientific process of moving from simulated to real world testing, and are taught the CPS core competencies of mathematics, physics, programming languages, and other tools. Additionally, students are introduced to the concepts of peer review, and iterative development. Student feedback showed that students greatly enjoy the program, and students indicate interest in future participation in CPS related research activities. We also give the reader the tools and resources required to recreate the summer educational program.

\section{Acknowledgments}
This work was supported by a National Science Foundation Grant. The authors would also like to thank Paul Sivilotti, Keith Redmill, Scott Biddlestone, Arda Kurt, and Micheal Vernier for their support in developing the education activities.

\bibliographystyle{abbrv}
\bibliography{josce_paper}  
\raggedbottom

\end{document}